\documentclass[journal]{IEEEtran}

\newtheorem{remark}{Remark}
\newtheorem{lemma}{Lemma}
\newtheorem{theorem}{Theorem}
\newtheorem{definition}{Definition}

\usepackage{amssymb}

\usepackage{cite}

\usepackage{graphicx}
\usepackage{subfigure}

\usepackage{amsmath}

\usepackage{algorithm}
\usepackage{algorithmic}
\usepackage{multirow}
\renewcommand{\algorithmicrequire}{\textbf{Input:}} 

\usepackage{float}

\begin{document}
%
\title{Multi-Agent Q-Learning Aided Backpressure Routing Algorithm for Delay Reduction}


\author{Juntao Gao,
				Yulong Shen,
				Minoru Ito,
        and Norio Shiratori
\thanks{This material has been submitted in part to International Conference on Computing, Networking and Communications (ICNC) 2018, Hawaii, USA.}
\thanks{J. Gao and M. Ito are with the Graduate School of Information Science, Nara Institute of Science and Technology, Nara 630-0192, Japan. E-mail: \{jtgao,ito\}@is.naist.jp.}
\thanks{Y. Shen is with the School of Computer Science and Technology, Xidian University, Shaanxi 710071, China. E-mail: ylshen@mail.xidian.edu.cn.}
\thanks{N. Shiratori is with Research Institute of Electrical Communication, Tohoku University, Sendai 980-8579, Japan.}}

\maketitle

\begin{abstract}
In queueing networks, it is well known that the throughput-optimal backpressure routing algorithm results in poor delay performance for light and moderate traffic loads. 
To improve delay performance, state-of-the-art backpressure routing algorithm (called BPmin \cite{Cui_TON16}) exploits queue length information to direct packets to less congested routes to their destinations.
However, BPmin algorithm estimates route congestion based on unrealistic assumption that every node in the network knows real-time global queue length information of all other nodes.
In this paper, we propose multi-agent Q-learning aided backpressure routing algorithm, where each node estimates route congestion using only local information of neighboring nodes.
Our algorithm not only outperforms state-of-the-art BPmin algorithm in delay performance but also retains the following appealing features: distributed implementation, low computation complexity and throughput-optimality.
Simulation results show our algorithm reduces average packet delay by $95\%$ for light traffic loads and by $41\%$ for moderate traffic loads when compared to state-of-the-art BPmin algorithm.


\end{abstract}

\IEEEpeerreviewmaketitle

\section{Introduction}
\begin{figure}[!th]
\centering
\includegraphics[width=3.1in]{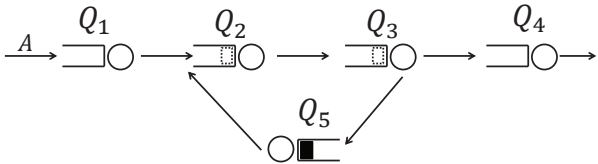}
\caption{A packet directed to a route loop, experiencing long delay.}
\label{fig:routeloop}
\end{figure}

Backpressure routing algorithm, which routes packets in a queueing network by congestion gradients, holds great potentials for applications in different areas, like sensor networks \cite{Jiao_JWCN15}, mobile ad hoc networks \cite{Georgiadis_FTN06} and transportation systems \cite{Gregoire_CNS15,Zaidi_ITS16}.
It is well known that the backpressure routing algorithm achieves maximum network throughput (\textit{throughput optimality}) by exploring all possible routes (even route loops) to balance traffic loads over the entire queueing network. 
This is effective for queueing networks with heavy traffic loads. 
However, for light and moderate traffic loads, excessive route exploration may lead to packets being directed to unnecessarily long routes or even route loops as shown in Fig. \ref{fig:routeloop},
which results in poor delay performance \cite{Cui_IT12}. 

To improve delay performance of backpressure routing algorithm, available works \cite{Yin_ANC17,Cui_TON16,Bui_TON13,Neely_JSAC05,Ying_TON11,Jiao_JWCN15,Ji_TON13} aim at directing packets to shorter routes to their destinations by exploiting various information of queueing networks, such as information of queue length, shortest path length (distance of the shortest path between two nodes) and packet delay (see Section \ref{section:rw} for details). Out of these works, state-of-the-art BPmin algorithm proposed in \cite{Cui_TON16} significantly reduces average packet delay of backpressure routing algorithm. According to BPmin algorithm, every node in a queueing network needs to know queue length information of all other nodes in real time. Based on these queue length information, every node calculates the sum of queue length of each route as route congestion estimate and then directs packets to least congested routes to their destinations. However, in queueing networks it is unrealistic for nodes to collect such real-time global queue length information. Moreover, BPmin algorithm requires the knowledge of network throughput capacity to make routing decisions, which is hard to determine.

In this paper, we propose multi-agent Q-learning aided backpressure routing algorithm (QL-BP), where each node estimates route congestion using only local information of neighboring nodes.
Specifically, every node under our QL-BP algorithm maintains multiple Q-learning agents, where each Q-learning agent continuously updates its route congestion estimate using neighboring nodes' queue length information and neighboring nodes' route congestion estimates. 
Based on estimated route congestion, every node directs packets to least congested routes to their destinations.
Our algorithm not only outperforms state-of-the-art BPmin algorithm in delay performance but also retains the following appealing features: distributed implementation, low computation complexity and throughput-optimality.
Simulation results show our algorithm reduces average packet delay by $95\%$ for light traffic loads and by $41\%$ for moderate traffic loads when compared to state-of-the-art BPmin algorithm.

The rest of this paper is organized as follows. In Section \ref{section:nm}, we introduce network models concerning communication links, resource allocation, transmission rates, packet generating, etc. In Section \ref{section:gf}, we describe in details our multi-agent Q-learning aided backpressure routing algorithm (QL-BP). In Section \ref{section:pa}, we analyze the performance of our QL-BP algorithm. In Section \ref{section:simu}, we do simulations to evaluate the delay performance of our QL-BP algorithm. We introduce related work in Section \ref{section:rw} and conclude the whole paper in Section \ref{section:conclusion}.


\section{Network Model}  \label{section:nm}

\begin{table*}[!htbp]
\caption{key notations}
\label{table:notation}
\centering
\begin{tabular}{ |l| l| }
\hline
\textbf{Notation} & \textbf{Definition} \\ \hline
$(i,j)$ & The communication link between two nodes $i$ and $j$, $i,j \in \mathcal{N}$, is denoted by pair $(i,j)$, $(i,j)\in \mathcal{L}$, \\
        & which is different from link $(j,i)$. \\ \hline
$S_{ij}(t)$ & The state of link $(i,j)$ at slot $t$, which represents factors affecting transmission rate of link $(i,j)$ at slot $t$, \\
						& like node position, channel fading and interference coefficients. \\ \hline
$\boldsymbol{S}(t)$ & $\boldsymbol{S}(t)=\big(S_{ij}(t)\big)$, the matrix of all link states at slot $t$. \\ \hline
$I_{ij}(t)$ & The resource allocation decision over link $(i,j)$ at slot $t$, such as link activation, coding, modulation, etc.  \\ \hline
$\boldsymbol{I}(t)$ & $\boldsymbol{I}(t)=\big(I_{ij}(t)\big)$, the matrix of resource allocation decision over all links at slot $t$. \\ 
										& $\boldsymbol{I}(t)$ (respectively, $I_{ij}(t)$) under algorithm $X$ is denoted by $\boldsymbol{I}^{X}(t)$ (resp. $I^{X}_{ij}(t)$). \\ \hline
$\mu_{ij}(t)$  & The offered transmission rate (packets/slot) over link $(i,j)$ at slot $t$ under $\boldsymbol{S}(t)$ and $\boldsymbol{I}(t)$. \\
							&  Actual data amount transmitted over link $(i,j)$ during slot $t$ may be less than $\mu_{ij}(t)$ due to insufficient data. \\ \hline
$\boldsymbol{\mu}(t)$ & $\boldsymbol{\mu}(t)=\big( \mu_{ij}(t)\big)$, the matrix of offered transmission rates over all links. \\ \hline
$\mu_{ij}^{(c)}(t)$ & The offered transmission rate to commodity $c$ over link $(i,j)$ at slot $t$, $\sum_c \mu_{ij}^{(c)}(t) \leq \mu_{ij}(t)$. \\
										& Actual data amount of commodity $c$ transmitted over link $(i,j)$ during slot $t$ may be less than $\mu_{ij}^{(c)}(t)$.\\ 
										& Further, $\mu_{ij}^{(c)}(t)$ under algorithm $X$ is denoted by $\mu_{ij}^{(c)X}(t)$. \\ \hline
$A_i^{(c)}(t)$ & The amount of packets node $i$ generates at slot $t$, which are destined for node $c, c \neq i$. \\ \hline
$U_i^{(c)}$ & The queue of node $i$, which stores packets destined for node $c \in \{1, 2, \cdots, N\}, c\neq i$. \\ \hline
$U_i^{(c)}(t)$ & The queue length of queue $U_i^{(c)}$ at slot $t$, i.e, the number of packets queueing up at queue $U_i^{(c)}$ at slot $t$. \\
							&  By convention, $U_i^{(i)}(t)=0$. \\ \hline
$\boldsymbol{U}(t)$ & $\boldsymbol{U}(t)=\big(U_i^{(c)}(t)\big)$, the matrix of queue length of all queues at slot $t$. \\ \hline
$B_i^{(c)}(t)$ & The bias associated with queue $U_i^{(c)}$ at slot $t$. \\ \hline
$\boldsymbol{B}(t)$ & $\boldsymbol{B}(t)=\big(B_i^{(c)}(t)\big)$, the matrix of bias for all queues at slot $t$. \\ \hline
$\boldsymbol{H}(t)$ & Matrix of information of a queueing network, which is used to extract bias.  \\ 
										& Examples include information of queue length, shortest path length, packet delay, etc.\\ \hline
$Q_{ij}^{(c)}$ & Route congestion estimated by Q-learning agent of node $i$ for routes of commodity $c$ and by the way of \\
									&  node $i$'s neighbor $j$. \\ \hline
$\boldsymbol{Q}_i$ & $\boldsymbol{Q}_i=(Q_{ij}^{(c)})$, the matrix of route congestion estimates of node $i$. \\ \hline 
\end{tabular}
\end{table*}

\begin{figure}[!th]
\centering
\includegraphics[width=2.6in]{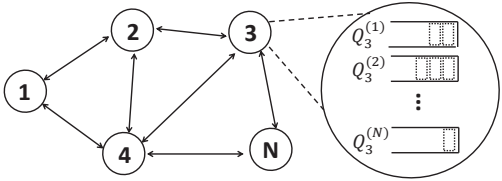}
\caption{Multi-hop queueing network and queues of node $3$.}
\label{fig:topology}
\end{figure}

We consider a multi-hop queueing network represented by a directed graph $\mathcal{G}=(\mathcal{N},\mathcal{L})$ as shown in Fig. \ref{fig:topology}, where $\mathcal{N}$ is the set of $N$ nodes and $\mathcal{L}$ is the set of $L$ directed links. The whole network operates over discrete time slots $t \in \{0, 1, 2, \cdots\}$. In the network, every node both transmits packets generated by itself and relays packets from other nodes to their destinations. For this purpose, each node maintains seperate queues to store packets destined for different destinations. For example, queue $U_i^{(c)}$ of node $i$ stores packets destined for node $c \in \{1, 2, \cdots, N\}, c\neq i$. All packets destined for the same destination $c$ are referred to as commodity $c$. Key definitions and notations to be used in the following are summarized in Table \ref{table:notation}.

Transmission rate $\mu_{ij}(t)$ is affected by random link states $\boldsymbol{S}(t)$ and resource allocation decisions $\boldsymbol{I}(t)$ such that 
\begin{align}
\mu_{ij}(t) & = \hat{\mu}_{ij}(\boldsymbol{S}(t),\boldsymbol{I}(t)), \quad \boldsymbol{S}(t) \in \mathcal{S}, \boldsymbol{I}(t) \in \mathcal{I}
\end{align} 
where $\mathcal{S}$ is the finite space of link states and $\mathcal{I}$ is the finite space of resource allocation decisions.
We assume that the outgoing transmission rate and incoming transmission rate of all nodes are upper bounded
\begin{align}
\mu_{max}^{out} & = \max_{i,s \in \mathcal{S}, I \in \mathcal{I}} \sum_j \hat{\mu}_{ij}(s,I) \label{equation:outmax} \\
\mu_{max}^{in} & = \max_{i,s \in \mathcal{S}, I \in \mathcal{I}} \sum_k \hat{\mu}_{ki}(s,I) \label{equation:inmax}
\end{align} 
The amount of packets generated by each node at each slot $t$ is also upper bounded by a positive constant $A_{max}$ such that
\begin{align}
\max_i \mathbb{E} \Big\{ \Big[ \sum_c A_i^{(c)}(t)\Big]^2 \Big\} & \leq A_{max}^2 \label{equation:Amax}
\end{align}

After $K$ time slots (called convergence interval in \cite{Neely_PhD03}), the queueing network arrives at steady state such that packet generating processes $A_i^{(c)}(t)$ and link states $\boldsymbol{S}(t)$ converge as follows
\begin{align}
\Big|\frac{1}{K}\sum_{\tau=t_0}^{t_0+K-1}\mathbb{E}\big\{A_i^{(c)}(t)\big\}-\lambda_i^{(c)}\Big| & \leq \delta_1 \label{equation:AK} \\
\sum_{s \in \mathcal{S}}\Big|\frac{1}{K}\sum_{\tau=t_0}^{t_0+K-1}\mathbb{E}\big\{\boldsymbol{1}_{[\boldsymbol{S}(t)=s]}\big\}-\pi_s\Big| & \leq \delta_2  \label{equation:SK}
\end{align}
where $\boldsymbol{1}_{[statement]}$ is an indicator function that returns value $1$ if $statement$ is true and $0$ otherwise,  $\lambda_i^{(c)}$ is the average rate of packet generating process $A_i^{(c)}(t)$, $\pi_s$ is the rate of link states $\boldsymbol{S}(t)$ being at state $s\in \mathcal{S}$, $\delta_1$ and $\delta_2$ are two small positive numbers. 

Queue length $U_i^{(c)}(t)$ for two adjacent slots satisfies the following relationship
\begin{align} \label{equation:aqd}
U_i^{(c)}(t+1)&\leq \max\Big\{U_i^{(c)}(t) - \sum_j \mu_{ij}^{(c)}(t), 0\Big\} \nonumber \\
              & \quad \quad +\sum_k \mu_{ki}^{(c)}(t)+A_i^{(c)}(t)
\end{align}
because $\mu_{ki}^{(c)}(t)$ is the offered transmission rate to commodity $c$ over link $(k,i)$ at slot $t$, however, node $k$ may not have enough packets to transmit to node $i$ at slot $t$.
 
\section{Multi-Agent Q-Learning Aided Backpressure Routing Algorithm} \label{section:gf}
In this section we introduce in details our multi-agent Q-learning aided backpressure routing algorithm (QL-BP).
First, we propose a bias based general framework for delay reduction in backpressure routing algorithm. 
Then, we build QL-BP algorithm based on this general framework.

\subsection{Bias Based General Framework}
\begin{figure}[!th]
\centering
\includegraphics[width=3.3in]{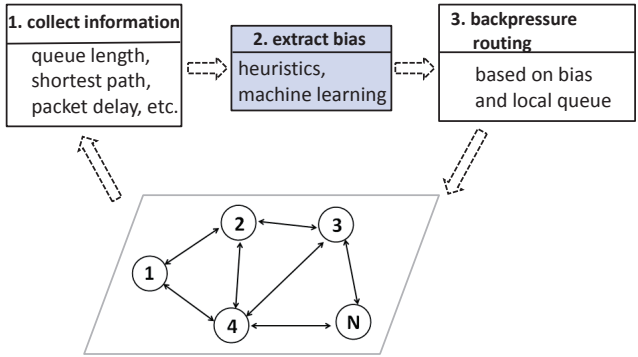}
\caption{Bias based framework in backpressure routing algorithm.}
\label{fig:bpbias}
\end{figure}


\begin{algorithm}[!htbp]
\caption{Bias Based General Framework for Delay Reduction in Backpressure Routing (BPBias)}
\label{algorithm:BPBias}
\begin{algorithmic}[1]
\renewcommand{\algorithmicrequire}{\underline{\textbf{Information collection:}}}
\REQUIRE  At every time slot $t$, network controller observes link states $\boldsymbol{S}(t)$ and collects information $\boldsymbol{H}(t)$, such as information of (local or global) queue length, shortest path for all node pairs, packet delay for all queues.
\renewcommand{\algorithmicrequire}{\underline{\textbf{Bias extraction:}}}
\REQUIRE Network controller extracts bias $\boldsymbol{B}(t)$ from $\boldsymbol{H}(t)$ for delay reduction. Various bias extracting methods can be adopted here, such as heuristic and machine learning methods.
\renewcommand{\algorithmicrequire}{\underline{\textbf{Bias Based Backpressure routing:}}}
\REQUIRE
\STATE For all links $(i,j)$, find the optimal commodity $c_{ij}^{\ast}(t)$ such that
			\begin{align}
				c_{ij}^{\ast}(t)\!=\!\underset{c\in \mathcal{N}}{\mathrm{argmax}} \Big\{\!\!\Big(U_i^{(c)}\!(t)\!+\!B_i^{(c)}\!(t)\Big)\!\!-\!\Big(U_j^{(c)}\!(t)\!+\!B_j^{(c)}\!(t)\Big)\!\Big\}
			\end{align}
			Calculate pressure gradients $W_{ij}^{\ast}(t)$ for all links $(i,j)$
			\begin{align} \label{equation:weight}
				W_{ij}^{\ast}(t)&= \max{\Big\{\!\Big(U_i^{(c_{ij}^{\ast}(t))}\!(t)\!+\!B_i^{(c_{ij}^{\ast}(t))}\!(t)\Big)} \nonumber \\
				                & \quad\quad\quad\quad -\!\Big(U_j^{(c_{ij}^{\ast}(t))}\!(t)\!+\!B_j^{(c_{ij}^{\ast}(t))}\!(t)\Big)\!,0\Big\}
			\end{align}
			Make resource allocation decision $\boldsymbol{I}^{BPBias}(t)$ such that
			\begin{align} \label{equation:ra}
				\boldsymbol{I}^{BPBias}(t)= \underset{I \in \mathcal{I}}{\mathrm{argmax}} \sum_{(i,j)\in \mathcal{L}} \hat{\mu}_{ij}\big(\boldsymbol{S}(t),I\big)\cdot W_{ij}^{\ast}(t)
			\end{align}
\STATE Under $\boldsymbol{I}^{BPBias}(t)$ and $\boldsymbol{S}(t)$, for each link $(i,j)$ and commodity $c$, transmit packets with rates as follows
\begin{align}  \label{equation:BPBiasrate}
\mu_{ij}^{(c)BPBias}(t) \!=\!
					\left\{
							\begin{array}{l l}
								\hat{\mu}_{ij}\big(\boldsymbol{S}(t),\boldsymbol{I}^{BPBias}(t)\big)&  \text{if $c=c_{ij}^{\ast}(t)$,} \\
								  & \text{$W_{ij}^{\ast}(t)\!>\!0$;}\\
								0&  \text{otherwise.}
					    \end{array} 
					\right.
\end{align}
\end{algorithmic}
\end{algorithm}

The whole bias based general framework consists of three stages: information collection, bias extraction and backpressure routing, as illustrated in Fig. \ref{fig:bpbias} and summarized in Algorithm \ref{algorithm:BPBias}.
At stage of information collection, our framework (referred to as BPBias) collects useful (local or global) information $\boldsymbol{H}(t)$, like queue length, shortest path and packet delay, for delay reduction.
At stage of bias extraction, BPBias framework extracts useful features (e.g., route congestion estimate) from $\boldsymbol{H}(t)$ as matrix of bias $\boldsymbol{B}(t)=(B_i^{(c)}(t))$, where $B_i^{(c)}(t)$ is the bias for queue $U_i^{(c)}$ at slot $t$ and is a function of $\boldsymbol{H}(t)$ and upper bounded by a positive constant $B_{max}$,
\begin{align}
&B_i^{(c)}(t) = B_i^{(c)}(\boldsymbol{H}(t)) \\
&0 \leq B_i^{(c)}(t) \leq B_{max}, \quad \text{for all $i,c$}  \label{inequality:biasBound} \\
& B_i^{(i)}(t)=0, \quad \text{for all $i$}
\end{align} 
At the final stage of backpressure routing, BPBias framework programs extracted bias $\boldsymbol{B}(t)$ into backpressuring routing algorithm, enabling the algorithm to adaptively change packet routes for delay reduction. Specifically, bias $B_i^{(c)}(t)$ for different queue $U_i^{(c)}$ can be dynamically adjusted according to real-time information $\boldsymbol{H}(t)$, so that packets can be directed to better routes. 



The methods for extracting bias $\boldsymbol{B}(t)$ from $\boldsymbol{H}(t)$ can be either heuristic based methods \cite{Yin_ANC17,Cui_TON16,Bui_TON13,Neely_JSAC05,Ying_TON11,Jiao_JWCN15,Ji_TON13} or machine learning based methods, like Q-learning \cite{Sutton_RL98}. This flexibility enables our framework to be very general and cover many bias based backpressure routing algorithms as special cases as listed in Table \ref{table:specialCases}. 

\begin{table}[ht]
\caption{Special Cases of Our Framework}
\label{table:specialCases}
\centering
\begin{tabular}{ |l| l| }
\hline
\textbf{Special Case} & \textbf{Condition} \\ \hline
\cite{Yin_ANC17}    & $B_i^{(c)}(t)$ varies with real-time queue length \\
										& and packet delay \\\hline
\cite{Cui_TON16}    & $B_i^{(c)}(t)$ is set to be bias functions in \cite{Cui_TON16} \\\hline
\cite{Bui_TON13}    & $B_i^{(c)}(t)-B_j^{(c)}(t)=-M$ for all links $(i,j)$ \\ \hline
\cite{Neely_JSAC05} & $B_i^{(c)}(t)$ is the shortest distance \\ 
										& between node $i$ and node $c$ \\ \hline
\cite{Ying_TON11}   & $B_i^{(c)}(t)$ contains information of constrains \\
										& on route length \\ \hline
\cite{Jiao_JWCN15}  & $B_i^{(c)}(t)$ is a constant calculated as in \cite{Jiao_JWCN15} \\ \hline
\cite{Ji_TON13}			& $B_i^{(c)}(t)$ is a function of packet delay information \\ \hline
\end{tabular}
\end{table}


\subsection{QL-BP Algorithm}
Based on this general BPBias framework, we propose multi-agent Q-learning aided backpressure routing algorithm (QL-BP), where Q-learning agents are responsible for extracting route congestion estimate from collected information, which is used as bias to aid backpressure routing algorithm to reduce packet delay. 

From now on, we focus on queueing networks with independent links (e.g., wireline networks or wireless networks with orthogonal links).
Under our QL-BP algorithm, each node $i$ maintains multiple Q-learning agents, where each agent $QLA_{ij}^{(c)}$ is associated with one commodity $c$ and one neighbor $j$ of node $i$, responsible for estimating the route congestion $Q_{ij}^{(c)}$ for routes of commodity $c$ and by the way of node $i$'s neighbor $j$. Thus, each node $i$ maintains a table $\boldsymbol{Q}_i=(Q_{ij}^{(c)})$ storing route congestion estimates.

At stage of information collection, each node $i$ observes local link states $S_{ij}(t)$ and collects local information by exchanging its own queue length $U_i^{(c)}(t)$ and table $\boldsymbol{Q}_i$ of route congestion estimates with its neighboring nodes. At stage of bias extraction, each Q-learning agent $QLA_{ij}^{(c)}$ of node $i$ updates its route congestion estimate $Q_{ij}^{(c)}$ as follows:
\begin{align}
Q_{ij}^{(c)} \leftarrow (1-\alpha)Q_{ij}^{(c)}+\alpha \big[U_j^{(c)}(t)+\gamma \min_{k}Q_{jk}^{(c)}\big]
\end{align}
where $\alpha$ and $\gamma$ are Q-learning parameters, $0<\alpha, \gamma \leq 1$. If $Q_{ij}^{(c)}>B_{max}$, set $Q_{ij}^{(c)}=B_{max}$.
Each node $i$ calculates bias $B_i^{(c)}(t)$ for commodity $c$ as 
\begin{align}
B_i^{(c)}(t)=\min_j Q_{ij}^{(c)} 
\end{align}
Finally, at stage of backpressure routing, based on extracted bias $B_i^{(c)}(t)$ and observed link states $S_{ij}(t)$ each node $i$ makes resource allocation and routing decisions as in BPBias.

From the description of QL-BP algorithm, we see that QL-BP algorithm is a special case of BPBias framework. For queueing networks with independent links, transmission rates $\mu_{ij}(t)$ of all links are also independent of each other. Thus, the maximum of the weighted sum of (\ref{equation:ra}) can be achieved by each node independently maximizing corresponding terms as follows:
\begin{align} \label{equation:ra1}
&\underset{I \in \mathcal{I}}{\mathrm{max}} \sum_{(i,j)\in \mathcal{L}} \hat{\mu}_{ij}\big(\boldsymbol{S}(t),I\big)\cdot W_{ij}^{\ast}(t) \\ \nonumber
=&\sum_i \Big[\underset{I_{ij} }{\mathrm{max}} \sum_{j:(i,j)\in \mathcal{L}} \hat{\mu}_{ij}\big(S_{ij}(t),I_{ij} \big)\cdot W_{ij}^{\ast}(t)\Big]
\end{align}
where $I_{ij}$ denotes the available resource allocation decision for link $(i,j)$.
Therefore, our QL-BP algorithm can be implemented in a distributed way. 
Furthermore, each node under QL-BP algorithm only needs to exchange information with neighboring nodes and maximizes weighted sum locally, the computation complexity of QL-BP algorithm is low as compared to algorithms globally maximizing the weighted sum of (\ref{equation:ra}).

\remark Our QL-BP algorithm can be further improved by considering shortest path information (referred to as QLSP-BP). 
Under QLSP-BP algorithm, each node $i$ calculates bias for commodity $c$ as 
\begin{align}
B_i^{(c)}(t)=\min_j Q_{ij}^{(c)} + P_i^{(c)}
\end{align}
where $P_i^{(c)}$ is the length of the shortest path from node $i$ to node $c$.
The rest of QLSP-BP is the same with QL-BP.

\section{Algorithm Performance Analysis} \label{section:pa}
In this section, we show that our QL-BP algorithm is also throughput-optimal.

First, we introduce the following definitions concerning queueing network stability, queueing network stability region and throughput optimality.

\begin{definition}[Network Stability \cite{Neely_PhD03}]
A single queue $U_i^{(c)}$ is said to be stable if $g(V)\rightarrow 0$ as $V \rightarrow \infty$, where
\begin{align}
g(V)=\limsup_{t \rightarrow \infty} \mathbb{E}\Big\{\frac{1}{t}\int_0^t \boldsymbol{1}_{[U_i^{(c)}(t)>V]}dt \Big\}
\end{align}
A queueing network is said to be stable if all queues are stable.
\end{definition}
\begin{definition}[Network Stability Region]
Packet generating rates $\big(\lambda_i^{(c)}\big)$ are said to be supported by a queueing network if the queueing network can be stabilized by some routing algorithm under $\big(\lambda_i^{(c)}\big)$. 
The network stability region $\Lambda$ is the closure of the set of all packet generating rates $\big(\lambda_i^{(c)}\big)$ that can be supported by the queueing network.
\end{definition}
\begin{definition}[Throughput Optimality]
An algorithm is said to be throughput optimal if it can stabilize the queueing network for all packet generating rates $\big(\lambda_i^{(c)}\big)$ that are within network stability region $\Lambda$, i.e., $\big(\lambda_i^{(c)}+\epsilon\big) \in \Lambda$, $\epsilon>0$.
\end{definition}

Then, we establish the throughput-optimality of our QL-BP algorithm.
\begin{theorem}[QL-BP Throughput-Optimality]
For a queueing network $\mathcal{G}=(\mathcal{N},\mathcal{L})$ with stability region $\Lambda$, our QL-BP algorithm is throughput optimal.
\end{theorem}

\textit{Proof:} 
Since QL-BP algorithm is a special case of BPBias framework, if we can prove that the general BPBias is throughput optimal, then QL-BP algorithm is also throughput optimal.

According to the definition of throughput-optimality, we need to prove that for any packet generating rates $\big(\lambda_i^{(c)}\big)$ within network stability region $\Lambda$, i.e., $\big(\lambda_i^{(c)}+\epsilon\big) \in \Lambda$, $\epsilon>0$, our framework BPBias can stabilize the queueing network. Some steps of the following proof are similar to that of \cite{Neely_PhD03}, which are included here for completeness.

Recall that $K$ is the convergence interval of queueing network $\mathcal{G}$.
For any routing algorithm and time interval $[t_0, t_0+K-1]$, queue length $U_i^{(c)}(t_0+K), i\neq c,$ satisfies the following relationship
\begin{align} \label{equation:qd}
U_i^{(c)}(t_0+K)&\leq \max\Big\{U_i^{(c)}(t_0) - K\sum_j \overline{\mu}_{ij}^{(c)}, 0\Big\} \nonumber \\
              & \quad \quad +K\sum_k \overline{\mu}_{ki}^{(c)}+K\overline{A}_{i}^{(c)}
\end{align}
where
\begin{align}
\overline{\mu}_{ij}^{(c)} & = \frac{1}{K}\sum_{\tau=t_0}^{t_0+K-1}\mu_{ij}^{(c)}(\tau) \\
\overline{A}_{i}^{(c)} & = \frac{1}{K}\sum_{\tau=t_0}^{t_0+K-1}A_i^{(c)}(\tau)
\end{align}
Refer to Appendix \ref{appedix:inequality1} for its derivation.
Since bias $B_i^{(c)}(t_0) \geq 0$ from $(\ref{inequality:biasBound})$, we have
\begin{align} \label{equation:KBqd}
U_i^{(c)}(t_0+K)&\leq \max\Big\{U_i^{(c)}(t_0)+B_i^{(c)}(t_0) - K\sum_j \overline{\mu}_{ij}^{(c)}, 0\Big\} \nonumber \\
              & \quad \quad +K\sum_k \overline{\mu}_{ki}^{(c)}+K\overline{A}_{i}^{(c)}
\end{align}
After squaring both sides of $(\ref{equation:KBqd})$ and basic algebraic manipulations, we get
\begin{align} \label{inequality:KBqdExpand}
&\big[U_i^{(c)}(t_0+K)\big]^2-\big[U_i^{(c)}(t_0)\big]^2 \nonumber \\
&\leq \big[ B_i^{(c)}(t_0)\big]^2 + K^2 \bigg[ \Big(\sum_j \overline{\mu}_{ij}^{(c)}\Big)^2+\Big(\sum_k \overline{\mu}_{ki}^{(c)}\Big)^2 \nonumber \\
& \quad +2\Big(\sum_k \overline{\mu}_{ki}^{(c)}\Big)\overline{A}_{i}^{(c)}+\Big(\overline{A}_{i}^{(c)}\Big)^2\bigg] + 2U_i^{(c)}(t_0)B_i^{(c)}(t_0) \nonumber \\
& \quad - 2K\Big[U_i^{(c)}(t_0)+B_i^{(c)}(t_0) \Big]\Big[ \sum_j \overline{\mu}_{ij}^{(c)}-\sum_k \overline{\mu}_{ki}^{(c)}-\overline{A}_{i}^{(c)}\Big]
\end{align}

Define Lyapunov function $L(\boldsymbol{U}(t))=\sum_{i\neq c}[U_i^{(c)}(t)]^2$. By summing $(\ref{inequality:KBqdExpand})$ over all nodes $i$ and commodities $c\neq i$ and taking conditional expectations, we get the $K$-step Lyapunov drift $\Delta\big(\boldsymbol{U}(t_0),\boldsymbol{B}(t_0)\big)$ as follows:
\begin{align}
&\Delta\big(\boldsymbol{U}(t_0),\boldsymbol{B}(t_0)\big) \nonumber \\
&=\mathbb{E}\Big\{L\big(\boldsymbol{U}(t_0+K)\big)-L\big(\boldsymbol{U}(t_0)\big)\Big|\boldsymbol{U}(t_0),\boldsymbol{B}(t_0)\Big\} \nonumber \\
& \leq C_1 +2\sum_{i \neq c}U_i^{(c)}(t_0)B_i^{(c)}(t_0)   \nonumber \\
& \quad  -2K\sum_{i\neq c}\Big[U_i^{(c)}(t_0)+B_i^{(c)}(t_0) \Big] \nonumber \\
& \quad \quad\quad \cdot\mathbb{E}\Big\{ \sum_j \overline{\mu}_{ij}^{(c)}-\sum_k \overline{\mu}_{ki}^{(c)}-\overline{A}_{i}^{(c)}\Big|\boldsymbol{U}(t_0),\boldsymbol{B}(t_0)\Big\} \label{inequality:lyapunovdrift}
\end{align}
where $C_1$ is a constant given by 
\begin{align}
&C_1  = K^2N\big[(\mu_{max}^{out})^2 +\big(\mu_{max}^{in}+A_{max})^2\big] \nonumber \\ 
& \quad\quad\quad + N(N-1)B_{max}^2 
\end{align}
and the expectation is with respect to random link states $\boldsymbol{S}(t)$, packet generating processes $A_i^{(c)}(t)$ and resource allocation decisions $\boldsymbol{I}(t)$.
Inequality $(\ref{inequality:lyapunovdrift})$ can also be written alternatively as 
\begin{align}
&\Delta\big(\boldsymbol{U}(t_0),\boldsymbol{B}(t_0)\big) \nonumber \\
&\leq C_1+2\sum_{i \neq c}U_i^{(c)}(t_0)B_i^{(c)}(t_0) \nonumber \\
& \quad -2K\Big[\Phi\big(\boldsymbol{U}(t_0),\boldsymbol{B}(t_0)\big)-\beta\big(\boldsymbol{U}(t_0),\boldsymbol{B}(t_0)\big)\Big]
\end{align}
where
\begin{align}
&\Phi\big(\boldsymbol{U}(t_0),\boldsymbol{B}(t_0)\big) =\frac{1}{K}\sum_{\tau=t_0}^{t_0+K-1}\mathbb{E}\Big\{\sum_{i\neq c}\Big[U_i^{(c)}(t_0)+B_i^{(c)}(t_0) \Big] \nonumber \\
& \quad\quad\quad\quad\quad\quad  \cdot\Big[\sum_j \mu_{ij}^{(c)}(\tau)-\sum_k \mu_{ki}^{(c)}(\tau)\Big]\Big|\boldsymbol{U}(t_0),\boldsymbol{B}(t_0)\Big\} \label{equation:Phi} \\
& \beta\big(\boldsymbol{U}(t_0),\boldsymbol{B}(t_0)\big) = \frac{1}{K}\sum_{\tau=t_0}^{t_0+K-1}\mathbb{E}\Big\{\sum_{i\neq c}\Big[U_i^{(c)}(t_0)+B_i^{(c)}(t_0) \Big] \nonumber \\
& \quad\quad\quad\quad\quad\quad  \cdot A_i^{(c)}(\tau)\Big|\boldsymbol{U}(t_0),\boldsymbol{B}(t_0)\Big\}
\end{align}

Next, we show that for any $\big(\lambda_i^{(c)}\big)$ such that $\big(\lambda_i^{(c)}+\epsilon\big) \in \Lambda$, $\epsilon>0$, our framework $BPBias$ stabilizes the queueing network and thus our framework is throughput-optimal.

Let $\Phi^{BPBias}\big(\boldsymbol{U}(t_0),\boldsymbol{B}(t_0)\big)$ and $\Phi^{X}\big(\boldsymbol{U}(t_0),\boldsymbol{B}(t_0)\big)$ be the quantity $\Phi\big(\boldsymbol{U}(t_0),\boldsymbol{B}(t_0)\big)$ under BPBias and under any other $X$ algorithm, respectively. Then, we have the following relationship
\begin{align} \label{inequality:BPBiasX}
\Phi^{BPBias}\big(\boldsymbol{U}(t_0),\boldsymbol{B}(t_0)\big) \geq \Phi^{X}\big(\boldsymbol{U}(t_0),\boldsymbol{B}(t_0)\big)-C_2
\end{align}
where $C_2$ is given in $(\ref{equation:C2})$. Refer to Appendix \ref{appedix:inequality1} for its derivation.

Thus, the $K$-step Lyapunov drift $\Delta\big(\boldsymbol{U}(t_0),\boldsymbol{B}(t_0)\big)$ under BPBias algorithm
\begin{align}
&\Delta\big(\boldsymbol{U}(t_0),\boldsymbol{B}(t_0)\big) \nonumber \\
&\leq C_1+2\sum_{i \neq c}U_i^{(c)}(t_0)B_i^{(c)}(t_0) \nonumber \\
& \quad -2K\Big[\Phi^{BPBias}\big(\boldsymbol{U}(t_0),\boldsymbol{B}(t_0)\big)-\beta\big(\boldsymbol{U}(t_0),\boldsymbol{B}(t_0)\big)\Big]  \\
& \leq C_1+2KC_2+2\sum_{i \neq c}U_i^{(c)}(t_0)B_i^{(c)}(t_0) \nonumber \\
& \quad -2K\Big[\Phi^{X}\big(\boldsymbol{U}(t_0),\boldsymbol{B}(t_0)\big)-\beta\big(\boldsymbol{U}(t_0),\boldsymbol{B}(t_0)\big)\Big]  \\
&= C_1+2KC_2+2\sum_{i \neq c}U_i^{(c)}(t_0)B_i^{(c)}(t_0) \nonumber \\
& -2K\sum_{i\neq c}\Big[U_i^{(c)}(t_0)+B_i^{(c)}(t_0) \Big]\frac{1}{K}\sum_{\tau=t_0}^{t_0+K-1}\mathbb{E}\bigg\{\sum_j \mu_{ij}^{(c)X}(\tau) \nonumber \\
& \quad -\sum_k \mu_{ki}^{(c)X}(\tau)-A_i^{(c)}(\tau)\Big|\boldsymbol{U}(t_0),\boldsymbol{B}(t_0)\bigg\} \label{inequality:Kdrift}
\end{align}
According to Theorem 6 \cite{Neely_PhD03}, we know that for any $\big(\lambda_i^{(c)}\big)$ such that $\big(\lambda_i^{(c)}+\epsilon\big) \in \Lambda$, $\epsilon>0$, there exists a stationary randomized routing algorithm $STAT$, which makes resource allocation and routing decisions independent of $\boldsymbol{U}(t_0)$ and $\boldsymbol{B}(t_0)$, such that 
\begin{align} \label{inequality:STAT}
&\frac{1}{K}\sum_{\tau=t_0}^{t_0+K-1}\mathbb{E}\bigg\{\sum_j \mu_{ij}^{(c)STAT}(\tau)-\sum_k \mu_{ki}^{(c)STAT}(\tau) \nonumber \\
&-A_i^{(c)}(\tau)\Big|\boldsymbol{U}(t_0),\boldsymbol{B}(t_0)\bigg\} \geq \frac{\epsilon}{2}
\end{align}
Substitute $(\ref{inequality:STAT})$ into $(\ref{inequality:Kdrift})$, we get 
\begin{align}
&\Delta\big(\boldsymbol{U}(t_0),\boldsymbol{B}(t_0)\big) \nonumber \\
&\leq C_1+2KC_2+2\sum_{i \neq c}U_i^{(c)}(t_0)B_i^{(c)}(t_0) \nonumber \\
&\quad-K\epsilon \sum_{i\neq c}\Big[U_i^{(c)}(t_0)+B_i^{(c)}(t_0) \Big]  \\
&\leq C_1+2KC_2+2\sum_{i \neq c}U_i^{(c)}(t_0)B_{max}-K\epsilon \sum_{i\neq c}U_i^{(c)}(t_0)  \\
&=C-(K\epsilon-2B_{max})\sum_{i\neq c}U_i^{(c)}(t_0) \label{inequality:Kdriftbound}
\end{align}
where $C=C_1+2KC_2$.

Since $K$ is the convergence interval of the queueing network, it is easy to determine the value of $K$ such that $K\epsilon-2B_{max}>0$, i.e., $K>2B_{max}/\epsilon$. From the $K$-step Lyapunov drift bound $(\ref{inequality:Kdriftbound})$ and Lemma 2 \cite{Neely_PhD03}, we know that our BPBias framework stabilizes the queueing network for any $\big(\lambda_i^{(c)}\big)$ such that $\big(\lambda_i^{(c)}+\epsilon\big) \in \Lambda$, $\epsilon>0$, and thus it is throughput optimal. Therefore, QL-BP algorithm is also throughput optimal. This completes the proof.

\section{Simulation} \label{section:simu}
In this section, we evaluate the delay performance of our QL-BP algorithm by simulations and compare it to other variants of backpressure routing algorithms.

\subsection{Simulation Setup}

\begin{figure}[!th]
\centering
\includegraphics[width=3.1in]{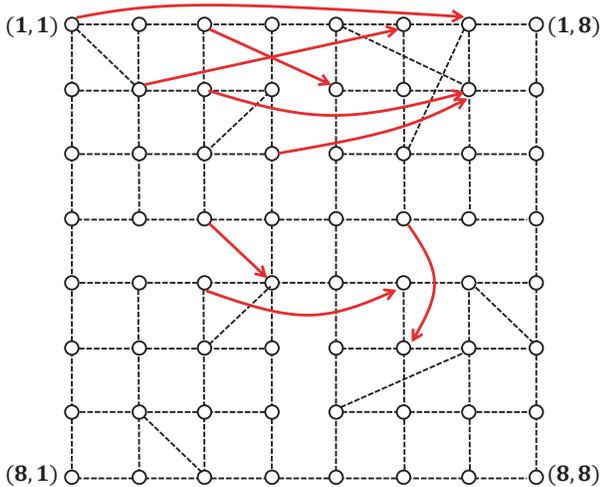}
\caption{Network topology for simulation.}
\label{fig:simu-topo}
\end{figure}

We consider the network topology as shown in Fig. \ref{fig:simu-topo}, which consists of $64$ nodes, indexed by a pair of coordinates. All links are bidirectional and the maximum data transmission rates for all links are 1 packet/slot. We assume all links can transmit packets simultaneously without interfering with each other, such as wireline network or wireless network with orthogonal channels. We consider traffic flows with the following source-destination pairs: ((1,3),(2,5)), ((2,3),(2,7)), ((2,2),(1,6)), ((3,4),(2,7)), ((1,1),(1,7)), ((4,3),(5,4)), ((4,6),(6,6)), and ((5,3),(5,6)). All source nodes generate packets according to Poisson distribution with rate $\lambda$ packets/slot. We implemented by Python our QL-BP algorithm, QLSP-BP algorithm, traditional backpressure routing algorithm (BP) \cite{Neely_JSAC05,Neely_PhD03}, shortest path based backpressure routing algorithm (SP-BP)\cite{Neely_JSAC05,Neely_PhD03} and state-of-the-art BPmin algorithm \cite{Cui_TON16}. For our QL-BP algorithm and QLSP-BP algorithm, we set Q-learning parameters $\lambda=1, \gamma=1$ to enable agents to quickly update their route congestion estimates. We run simulations for $10^5$ slots for each simulation setting and calculate the average delay of packets received by destinations under different algorithms. 

\subsection{Simulation Results}

\begin{figure}[!th]
\centering
\includegraphics[width=3.6in]{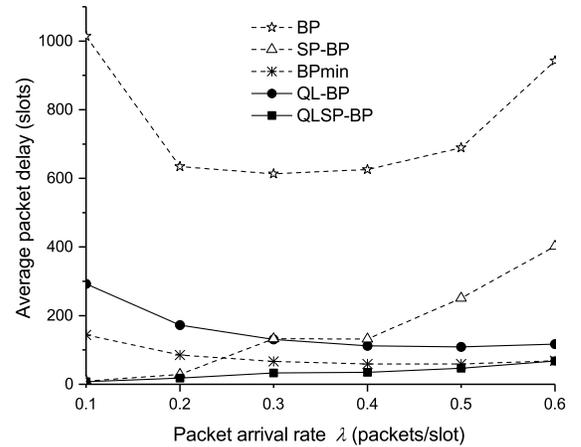}
\caption{Average packet delay under different backpressure routing algorithms.}
\label{fig:graph1}
\end{figure}

From Fig. \ref{fig:graph1}, we can observe that our QL-BP algorithm reduces average packet delay by $71\%$ when compared to traditional BP algorithm under light traffic loads with $\lambda=0.1$ and by $82\%$ under moderate traffic loads with $\lambda=0.4$, indicating that QL-BP algorithm effectively learns route congestion and adaptively directs packets to better routes. However, QL-BP algorithm results in higher packet delay than state-of-the-art BPmin algorithm. This is because nodes of BPmin algorithm know perfect real-time global queue length information and thus can accurately estimate congestion of different routes and direct packets to the least congested routes. Nodes of QL-BP algorithm only know local information of neighboring nodes, thus can only loosely estimate congestion of different routes, which may lead to directing packets to suboptimal routes. However, BPmin algorithm is not realistic since real-time global queue length information is hard to collect by nodes in real world. Our QL-BP algorithm trades off some packet delay for distributed algorithm implementation and low computation complexity, thus can be easily deployed in real queueing networks.

Our QL-BP algorithm can be greatly improved by considering shortest path information. From Fig. \ref{fig:graph1}, we see that QLSP-BP algorithm outperforms all variants of backpressure routing algorithms including state-of-the-art BPmin algorithm: reducing average packet delay by $95\%$ for light traffic loads with $\lambda=0.1$ and by $41\%$ for moderate traffic loads with $\lambda=0.4$ when compared to BPmin algorithm. In summary, our algorithm can be easily deployed in real queueing networks, but also achieves the best delay performance when compared to other variants of backpressure routing algorithms.

\section{Related Work} \label{section:rw}
The traditional backpressure routing algorithm routes packets according to congestion gradients, like water flowing through pipe networks according to pressure gradients \cite{Georgiadis_FTN06}. 
According to the traditional backpressure routing algorithm, the pressure of a queue is defined to be the number of packets queueing up at that queue (queue length). The pressure gradient between two queues of neighboring nodes is defined to be the difference of their queue pressure. 
The traditional backpressure routing algorithm routes packets based on only pressure gradients between neighboring nodes, i.e., local queue length information, without considering queue length of farther nodes and location of destinations. 
Its short-sightedness to farther nodes and blindness to destinations result in poor delay performance.

Available works on impoving delay performance of backpressure routing algorithm exploit various information of queueing networks, such as information of queue length \cite{Yin_ANC17,Cui_TON16,Bui_TON13}, shortest path length (distance of the shortest path between two nodes) \cite{Neely_JSAC05,Ying_TON11,Jiao_JWCN15} and packet delay \cite{Yin_ANC17,Ji_TON13}. 
Despite different forms of these works, they share the common characteristic: using bias to help backpressure routing to reduce packet delay.
According to whether bias value varies with time, these works are classified into two groups: backpressure routing with constant bias and backpressure routing with time-varying bias.

For backpressure routing with constant bias, Neely \textit{et al.} \cite{Neely_JSAC05} proposed an enhanced backpressure routing algorithm, where the constant bias is shortest path length. They combined the information of queue length and shortest path length to route packets in the direction of their destinations to shorten packet routes.
Ying \textit{et al.} \cite{Ying_TON11} also used shortest path length as constant bias, however in a different way, to shorten packet routes, where they imposed constraints on length of packet routes and maintained shortest path based queues to help meet constraints.
Instead of constructing constant bias from only shortest path length information, Jiao \textit{et al.} \cite{Jiao_JWCN15} built constant bias as a function of packet arrival rates, link transmission rates, and shortest path length.
Athanasopoulou \textit{et al.} \cite{Bui_TON13} proposed an $M$-Backpressure routing algorithm, where $M$ is a constant bias whose value is properly tuned to avoid long packet routes.
Yin \textit{et al.} \cite{Yin_ANC17} proposed a variant of backpressure routing algorithm, whose route searching process dynamically switches between shortest path mode and traditional backpressure routing mode based on constant bias (called threshold in \cite{Yin_ANC17}) to reduce packet delay.

For backpressure routing with time-varying bias, Ji \textit{et al.} \cite{Ji_TON13} introduced a delay-based backpressure routing algorithm to reduce packet delay for light traffic loads (called last packet problem in \cite{Ji_TON13}), where the time-varying bias is the delay of the head-of-line packet. 
Cui \textit{et al.} \cite{Cui_TON16} showed that time-varying bias based backpressure routing algorithms can significantly reduce packet delay. They also proposed two specific time-varying bias based backpressure routing algorithm: one considering local queue length information of up to two-hop nodes, the other considering global queue length information of all nodes (called BPmin). Out of these works, BPmin proposed in \cite{Cui_TON16} achieved state-of-the-art result in delay performance of backpressure routing algorithm.

\section{Conclusion} \label{section:conclusion}
In this paper we proposed multi-agent Q-learning aided backpressure routing algorithm. Our algorithm not only outperforms variants of backpressure routing algorithms, including state-of-the-art BPmin algorithm, in delay performance but also retains the following appealing features: distributed implementation, low computation complexity and throughput-optimality.
In the future, we will explore more advanced learning method (like deep learning \cite{Mnih_Nature15}) aided backpressure routing algorithm and do simulations to compare its delay performance with existing variants of backpressure routing.

\appendix

\subsection{Derivation of inequality $(\ref{equation:qd})$} 

We use mathematical induction to prove inequality $(\ref{equation:qd})$.
\underline{\textbf{Basis:}} From $(\ref{equation:aqd})$, we know for $l=1$
\begin{align} 
&U_i^{(c)}(t_0+l)=U_i^{(c)}(t_0+1) \nonumber \\
&\leq \max\Big\{U_i^{(c)}(t_0) - \sum_j \mu_{ij}^{(c)}(t_0), 0\Big\} \nonumber \\
& \quad\quad +\sum_k \mu_{ki}^{(c)}(t_0)+A_i^{(c)}(t_0)
\end{align}
\underline{\textbf{Inductive step:}} Assume the inequality holds for $l \geq 1$
\begin{align} \label{inequality:qd}
&U_i^{(c)}(t_0+l)  \nonumber \\
&\leq \max\Big\{U_i^{(c)}(t_0) - \sum_{\tau=t_0}^{t_0+l-1}\sum_j \mu_{ij}^{(c)}(\tau), 0\Big\} \nonumber \\ 
              & \quad \quad +\sum_{\tau=t_0}^{t_0+l-1}\sum_k \mu_{ki}^{(c)}(\tau)+\sum_{\tau=t_0}^{t_0+l-1}A_i^{(c)}(\tau)
\end{align}
We need to prove the inequality still holds for $l+1$.
From $(\ref{equation:aqd})$, we know that 
\begin{align} 
&U_i^{(c)}(t_0+l+1)\nonumber \\
&\leq \max\Big\{U_i^{(c)}(t_0+l) - \sum_j \mu_{ij}^{(c)}(t_0+l), 0\Big\} \nonumber \\
              & \quad \quad +\sum_k \mu_{ki}^{(c)}(t_0+l)+A_i^{(c)}(t_0+l)  \label{inequality:step1} \\
& \leq \max\bigg\{ \bigg[\max\Big\{U_i^{(c)}(t_0) - \sum_{\tau=t_0}^{t_0+l-1}\sum_j \mu_{ij}^{(c)}(\tau), 0\Big\} \nonumber \\
						& \quad \quad +\sum_{\tau=t_0}^{t_0+l-1}\sum_k \mu_{ki}^{(c)}(\tau)+\sum_{\tau=t_0}^{t_0+l-1}A_i^{(c)}(\tau)\bigg] \nonumber \\
						& \quad \quad - \sum_j \mu_{ij}^{(c)}(t_0+l), 0\bigg\}+\sum_k \mu_{ki}^{(c)}(t_0+l)+A_i^{(c)}(t_0+l)  \label{inequality:step2}\\
& \leq \max\bigg\{ \max\Big\{U_i^{(c)}(t_0) - \sum_{\tau=t_0}^{t_0+l-1}\sum_j \mu_{ij}^{(c)}(\tau), 0\Big\} \nonumber \\
						& \quad \quad- \sum_j \mu_{ij}^{(c)}(t_0+l), 0\bigg\}+\sum_{\tau=t_0}^{t_0+l}\sum_k \mu_{ki}^{(c)}(\tau)+\sum_{\tau=t_0}^{t_0+l}A_i^{(c)}(\tau) \label{inequality:step3}\\
&\leq \max\bigg\{ \max\Big\{U_i^{(c)}(t_0) - \sum_{\tau=t_0}^{t_0+l}\sum_j \mu_{ij}^{(c)}(\tau), 0\Big\}, 0\bigg\} \nonumber \\
& \quad \quad +\sum_{\tau=t_0}^{t_0+l}\sum_k \mu_{ki}^{(c)}(\tau)+\sum_{\tau=t_0}^{t_0+l}A_i^{(c)}(\tau) \label{inequality:step4}\\
&=\max\Big\{U_i^{(c)}(t_0) - \sum_{\tau=t_0}^{t_0+l}\sum_j \mu_{ij}^{(c)}(\tau), 0\Big\} \nonumber \\
& \quad \quad +\sum_{\tau=t_0}^{t_0+l}\sum_k \mu_{ki}^{(c)}(\tau)+\sum_{\tau=t_0}^{t_0+l}A_i^{(c)}(\tau) \label{inequality:step5}
\end{align}
where $(\ref{inequality:step2})$ follows from $(\ref{inequality:step1})$ and $(\ref{inequality:qd})$, $(\ref{inequality:step3})$ follows from $(\ref{inequality:step2})$ and $(\ref{equation:1})$, $(\ref{inequality:step4})$ follows from $(\ref{inequality:step3})$ and $(\ref{equation:2})$.
Thus, the inequality holds for any $l \geq 1$ including $l=K$.

\begin{lemma}
\begin{align}
\max \big[a, 0\big]+b  & \geq \max \big[a+b, 0\big],  \quad \text{for $b \geq 0$} \label{equation:1}\\
\max \big[a, 0\big]-b  & \leq \max \big[a-b, 0\big],  \quad \text{for $b \geq 0$} \label{equation:2}
\end{align}
\end{lemma}
\textbf{Proof:} For $b \geq 0$, we have
\begin{align} 
\max \big[a, 0\big]+b \!&=\!
					\left\{
							\begin{array}{l l}
								a+b&  \text{if $a \geq 0$}\\
								b&  \text{if $a<0$}
					    \end{array} 
					\right.  \label{equation:tmp1}\\ 
\max \big[a+b, 0\big] \!&=\!
					\left\{
							\begin{array}{l l}
								a+b&  \text{if $a \geq 0$}\\
								b-|a|&  \text{if $a<0, b-|a| \geq 0$} \\
								0 & \text{if $a<0, b-|a| <0$}
					    \end{array} 
					\right. \label{equation:tmp2}
\end{align}
By comparing equations $(\ref{equation:tmp1})$ with $(\ref{equation:tmp2})$, we have $(\ref{equation:1})$.
Similarly, we have
\begin{align} 
\max \big[a, 0\big]-b \!&=\!
					\left\{
							\begin{array}{l l}
								a-b&  \text{if $a \geq 0$}\\
								-b&  \text{if $a<0$}
					    \end{array} 
					\right.  \label{equation:tmp3}\\
\max \big[a-b, 0\big] \!&=\!
					\left\{
							\begin{array}{l l}
								a-b&  \text{if $a \geq 0, a-b \geq 0$}\\
								0&  \text{if $a \geq 0, a-b < 0$} \\
								0 & \text{if $a<0$}
					    \end{array} 
					\right. \label{equation:tmp4}
\end{align}
By comparing equations $(\ref{equation:tmp3})$ with $(\ref{equation:tmp4})$, we have $(\ref{equation:2})$.

\subsection{Derivation of inequality $(\ref{inequality:BPBiasX})$} \label{appedix:inequality1}

For the inner terms of $(\ref{equation:Phi})$, we have the following identity
\begin{align} \label{equation:identity}
&\sum_{i,c}\Big[U_i^{(c)}(t_0)+B_i^{(c)}(t_0) \Big]\Big[\sum_j \mu_{ij}^{(c)}(\tau)-\sum_k \mu_{ki}^{(c)}(\tau)\Big] \nonumber \\
& = \! \sum_{i,j}\!\sum_{c}\mu_{ij}^{(c)}(\tau)\Big[\Big(U_i^{(c)}(t_0)+B_i^{(c)}(t_0)\Big) \nonumber \\
&  \quad\quad\quad\quad\quad\quad\quad\quad - \Big(U_j^{(c)}(t_0)+B_j^{(c)}(t_0)\Big)  \Big]
\end{align}
where the condition $i\neq c$ is removed because $U_i^{(i)}(t)=0, B_i^{(i)}(t)=0$. Thus, quantity $\Phi\big(\boldsymbol{U}(t_0),\boldsymbol{B}(t_0)\big)$ can be alternatively written as 
\begin{align}
&\Phi\big(\boldsymbol{U}(t_0),\boldsymbol{B}(t_0)\big) \nonumber \\
& =\frac{1}{K}\sum_{\tau=t_0}^{t_0+K-1}\mathbb{E}\bigg\{\sum_{i,j}\!\sum_{c}\mu_{ij}^{(c)}(\tau)\Big[\Big(U_i^{(c)}(t_0)+B_i^{(c)}(t_0)\Big) \nonumber \\
& \quad \quad - \Big(U_j^{(c)}(t_0)+B_j^{(c)}(t_0)\Big)  \Big]\bigg|\boldsymbol{U}(t_0),\boldsymbol{B}(t_0) \bigg\}
\end{align}

For any routing algorithm $X$ and slot $\tau \in [t_0, t_0+K-1]$, we have the following relationship
\begin{align} 
&  \sum_{i,j}\!\sum_{c}\mu_{ij}^{(c)X}(\tau)\Big[\Big(U_i^{(c)}(\tau)+B_i^{(c)}(\tau)\Big) \nonumber \\
& \quad\quad\quad\quad\quad\quad\quad\quad- \Big(U_j^{(c)}(\tau)+B_j^{(c)}(\tau)\Big)  \Big] \nonumber \\
&  \leq \sum_{i,j}\!\sum_{c}\mu_{ij}^{(c)X}(\tau) \cdot W_{ij}^{\ast}(\tau) \label{inequality:XBPBias-step1} \\
& \leq \sum_{i,j}\!\hat{\mu}_{ij}\big(\boldsymbol{S}(\tau),\boldsymbol{I}^{X}(\tau)\big)\cdot W_{ij}^{\ast}(\tau) \label{inequality:XBPBias-step2} \\
& \leq \sum_{i,j}\!\hat{\mu}_{ij}\big(\boldsymbol{S}(\tau),\boldsymbol{I}^{BPBias}(\tau)\big)\cdot W_{ij}^{\ast}(\tau) \label{inequality:XBPBias-step3} \\
& = \sum_{i,j}\!\sum_{c}\mu_{ij}^{(c)BPBias}(\tau)\Big[\Big(U_i^{(c)}(\tau)+B_i^{(c)}(\tau)\Big) \nonumber \\
& \quad\quad\quad\quad\quad\quad\quad\quad\quad\quad - \Big(U_j^{(c)}(\tau)+B_j^{(c)}(\tau)\Big)  \Big] \label{inequality:XBPBias-step4}
\end{align}
where $(\ref{inequality:XBPBias-step1})$ follows from the definition of $W_{ij}^{\ast}(\tau)$ $(\ref{equation:weight})$, $(\ref{inequality:XBPBias-step2})$ follows from the fact that $\sum_{c}\mu_{ij}^{(c)X}(\tau) \leq \hat{\mu}_{ij}\big(\boldsymbol{S}(\tau),\boldsymbol{I}^{X}(\tau)\big)$, i.e., the sum of rates offered to all commodities over link $(i,j)$ is no greater than the total rate offered over link $(i,j)$, $(\ref{inequality:XBPBias-step3})$ follows from that BPBias algorithm achieves the maximum of weighted sum $(\ref{equation:ra})$, $(\ref{inequality:XBPBias-step4})$ follows from the definition of $\mu_{ij}^{(c)BPBias}(\tau)$ $(\ref{equation:BPBiasrate})$.

Using identity $(\ref{equation:identity})$, we write inequality $(\ref{inequality:XBPBias-step4})$ alternatively as 
\begin{align} \label{inequality:indentity2}
&\sum_{i,c}\Big[U_i^{(c)}(\tau)+B_i^{(c)}(\tau) \Big]\Big[\sum_j \mu_{ij}^{(c)BPBias}(\tau) \nonumber \\
&\quad\quad\quad\quad\quad\quad\quad\quad\quad\quad\quad\quad -\sum_k \mu_{ki}^{(c)BPBias}(\tau)\Big] \nonumber \\
&\geq \sum_{i,c}\Big[U_i^{(c)}(\tau)+B_i^{(c)}(\tau) \Big]\Big[\sum_j \mu_{ij}^{(c)X}(\tau)-\sum_k \mu_{ki}^{(c)X}(\tau)\Big]
\end{align}
Define $\Delta_i^{(c)}(\tau)=\big[U_i^{(c)}(\tau)+B_i^{(c)}(\tau) \big]-\big[U_i^{(c)}(t_0)+B_i^{(c)}(t_0) \big]$, then from $(\ref{inequality:indentity2})$ we have
\begin{align}
&\sum_{i,c}\Big[U_i^{(c)}(t_0)+B_i^{(c)}(t_0)+\big|\Delta_i^{(c)}(\tau)\big| \Big] \nonumber \\
&  \quad\quad \cdot\Big[\sum_j \mu_{ij}^{(c)BPBias}(\tau)-\sum_k \mu_{ki}^{(c)BPBias}(\tau)\Big] \nonumber \\
&\geq \sum_{i,c}\Big[U_i^{(c)}(t_0)+B_i^{(c)}(t_0)-\big|\Delta_i^{(c)}(\tau)\big| \Big] \nonumber \\
&  \quad\quad \cdot\Big[\sum_j \mu_{ij}^{(c)X}(\tau)-\sum_k \mu_{ki}^{(c)X}(\tau)\Big]
\end{align}
Then, we get
\begin{align} \label{inequality:compare}
&\sum_{i,c}\!\Big[U_i^{(c)}\!(t_0)\!+\!B_i^{(c)}\!(t_0) \Big] \nonumber \\
&  \quad\quad \cdot\Big[\sum_j \mu_{ij}^{(c)BPBias}(\tau)-\sum_k \mu_{ki}^{(c)BPBias}(\tau)\Big] \nonumber \\
& \quad\quad+ \sum_{i,c}\big|\Delta_i^{(c)}(\tau)\big|\Big[\mu_{max}^{out}+\mu_{max}^{in}\Big] \nonumber \\
&\geq \sum_{i,c}\Big[U_i^{(c)}(t_0)+B_i^{(c)}(t_0) \Big]\Big[\sum_j \mu_{ij}^{(c)X}(\tau)-\sum_k \mu_{ki}^{(c)X}(\tau)\Big] \nonumber \\
&  \quad\quad- \sum_{i,c}\big|\Delta_i^{(c)}(\tau)\big|\Big[\mu_{max}^{out}+\mu_{max}^{in}\Big]
\end{align}

Next, we show $\sum_{i,c}\big|\Delta_i^{(c)}(\tau)\big|$ is upper bounded. Note that
\begin{align} \label{inequality:upperbound0}
\big|\Delta_i^{(c)}(\tau)\big| \leq \big|U_i^{(c)}(\tau)-U_i^{(c)}(t_0)  \big|+\big|B_i^{(c)}(\tau)-B_i^{(c)}(t_0) \big|
\end{align}
From $(\ref{inequality:biasBound})$, we also have 
\begin{align}  \label{inequality:upperbound1}
0 \leq \big|B_i^{(c)}(\tau)-B_i^{(c)}(t_0) \big| \leq B_{max} 
\end{align}
Note that for any node $i$ and commodity $c$, the change of queue length from time slot $t_0$ to $\tau \in [t_0, t_0+K-1]$ is upper bounded as 
\begin{align} \label{inequality:upperbound2}
\big|U_i^{(c)}(\tau)\!-U_i^{(c)}(t_0)\big| \leq (\tau\!-\!t_0)(\mu_{max}^{out}+\mu_{max}^{in}\!+\!A_{max})
\end{align}


From $(\ref{inequality:upperbound0})$, $(\ref{inequality:upperbound1})$ and $(\ref{inequality:upperbound2})$, we know
\begin{align} \label{inequality:delta}
\sum_{i,c}\big|\Delta_i^{(c)}(\tau)\big| & \leq N(N-1)(\tau-t_0)(\mu_{max}^{out}+\mu_{max}^{in}+A_{max}) \nonumber \\
& \quad +N(N-1)B_{max}
\end{align}
Substitute $(\ref{inequality:delta})$ into $(\ref{inequality:compare})$, we get
\begin{align} \label{inequality:compare2}
&\sum_{i,c}\!\Big[U_i^{(c)}\!(t_0)\!+\!B_i^{(c)}\!(t_0) \Big] \nonumber \\
&  \quad\quad \cdot\Big[\sum_j \mu_{ij}^{(c)BPBias}(\tau)-\sum_k \mu_{ki}^{(c)BPBias}(\tau)\Big] \nonumber \\
&\geq \sum_{i,c}\Big[U_i^{(c)}(t_0)+B_i^{(c)}(t_0) \Big]\Big[\sum_j \mu_{ij}^{(c)X}(\tau)-\sum_k \mu_{ki}^{(c)X}(\tau)\Big] \nonumber \\
&  \quad\quad- 2\Big[N(N-1)(\tau-t_0)(\mu_{max}^{out}+\mu_{max}^{in}+A_{max}) \nonumber \\
&\quad\quad\quad\quad +N(N-1)B_{max}\Big]\Big[\mu_{max}^{out}+\mu_{max}^{in}\Big]
\end{align}
Take conditional expectations on both sides of $(\ref{inequality:compare2})$ and sum over $\tau \in [t_0, t_0+K-1]$, we have from $(\ref{equation:Phi})$ that 
\begin{align}
&\Phi^{BPBias}\big(\boldsymbol{U}(t_0),\boldsymbol{B}(t_0)\big) \nonumber \\
&\geq \Phi^{X}\big(\boldsymbol{U}(t_0),\boldsymbol{B}(t_0)\big) \nonumber \\
& \quad-\frac{2}{K}\sum_{\tau=t_0}^{t_0+K-1}\Big[N(N-1)(\tau-t_0)(\mu_{max}^{out}+\mu_{max}^{in}+A_{max}) \nonumber \\
& \quad\quad\quad \quad\quad\quad\quad +N(N-1)B_{max}\Big]\Big[\mu_{max}^{out}+\mu_{max}^{in}\Big]  \nonumber \\
& =\Phi^{X}\big(\boldsymbol{U}(t_0),\boldsymbol{B}(t_0)\big)-C_2
\end{align}
where
\begin{align} \label{equation:C2}
C_2 &=N(N-1)(K-1)(\mu_{max}^{out}+\mu_{max}^{in}+A_{max}) \nonumber \\
    & \quad \cdot(\mu_{max}^{out}+\mu_{max}^{in})+ 2N(N-1)B_{max}(\mu_{max}^{out}+\mu_{max}^{in})
\end{align}

\bibliographystyle{IEEEtran}
\bibliography{reference}

\end{document}